\documentclass[aip,jap,reprint,superscriptaddress]{revtex4-1}

\usepackage{amsmath,amssymb,amsthm,fancyhdr,units,mathtools, setspace, stmaryrd,graphicx,transparent,longtable,multirow}
\usepackage[english]{babel}
\setcitestyle{super}

\def\half{\frac{1}{2}}

\providecommand{\partder}[3]{\frac{\operatorname{\partial}^{#3}#1}{\partial#2^{#3}}} 			
\providecommand{\der}[3]{\frac{\operatorname{d}^{#3}\!#1}{\operatorname{d}\!#2^{#3}}} 		
\providecommand{\derop}[2]{\frac{\operatorname{d}^{#2}}{\operatorname{d}\!#1 ^{#2}}} 				
\providecommand{\on}[1]{\Big|_{#1}}
\providecommand{\exponential}[1]{\exp\left(#1\right)}

\begin{document}
\title{Numerical simulation of organic semiconductor devices with high carrier densities}

\author{S.~Stodtmann}
\email[]{sven.stodtmann@basf.com}
\affiliation{BASF SE, Scientific Computing Group, 67056 Ludwigshafen,
  Germany}
\affiliation{Interdisciplinary Center for Scientific Computing,
  University of Heidelberg, 69120 Heidelberg, Germany}

\author{R.~M.~Lee} 
\affiliation{BASF SE, Scientific Computing Group, 67056 Ludwigshafen,
  Germany}
\affiliation{Interdisciplinary Center for Scientific Computing,
  University of Heidelberg, 69120 Heidelberg, Germany}

\author{C.~K.~F.~Weiler}
\affiliation{Interdisciplinary Center for Scientific Computing,
  University of Heidelberg, 69120 Heidelberg, Germany}

\author{A.~Badinski}
\affiliation{BASF SE, Scientific Computing Group, 67056 Ludwigshafen,
  Germany}

\date{\today}

\begin{abstract}
  We give a full description of the numerical solution of a general
  charge transport model for doped disordered semiconductors with
  arbitrary field- and density-dependent mobilities. We propose a
  suitable scaling scheme and generalize the Gummel iterative
  procedure, giving both the discretization and linearization of the
  van Roosbroeck equations for the case when the generalized Einstein
  relation holds. We show that conventional iterations are unstable
  for problems with high doping, whereas the generalized scheme
  converges. The method also offers a significant increase in
  efficiency when the injection is large and reproduces known results
  where conventional methods converge.
\end{abstract}



\maketitle

\section{Introduction}
Organic semiconducting materials have been the subject of much
attention in recent years due to the prospect of tailoring their
photoelectric properties to specific
applications.\cite{Pron2010,Mensfoort2011,Anthony2010} 
 Doping organic semiconductors,
by embedding electron- or hole-donor species in the material, provides
a particularly effective way of tuning device
properties.\cite{Chan2009} Dopant concentrations can be easily and
precisely controlled, although the effect on device performance is not
always known \textit{a priori}.

Simulation is a powerful tool for understanding and optimizing device
performance, allowing systematic studies of device design that are
experimentally infeasible. Computational approaches taken by other
authors include Monte Carlo simulation,\cite{Holst2011} master
equation approaches,\cite{Mensfoort2010} and solution of the
one-dimensional drift-diffusion-Poisson system of
equations,\cite{Cottaar2012} which is the approach we follow
here. This last approach is highly suited to the extraction of model
parameters from experimental data and device optimization, since it is
less computationally-expensive than methods dealing with more
dimensions. One-dimensional drift-diffusion models originate from the
theory of conventional crystalline semiconductors. However, when
combining models for disordered materials with high doping
intensities, numerical approaches originally developed for inorganic
crystalline materials may fail. We derive a generalization of existing
methods, namely the Scharfetter-Gummel discretization\cite{Scharfetter1969} and the
Gummel-iteration map\cite{Gummel1964}, to overcome these problems and then compare our
results with a state-of-the-art simulation approach presented in
Ref.~\onlinecite{Knapp2010}. The numerical scheme that we develop here
improves the numerical stability and computational efficiency of the
approach.

The rest of this paper is structured as follows: In
Sec.~\ref{sec:model}, we introduce a model for steady-state charge
transport in disorded semiconductors. In Sec.~\ref{sec:scaling}, we
describe a scaling scheme to improve numerical behavior. We present in
Sec.~\ref{sec:discretization} the proper Scharfetter-Gummel
discretization of the model, accounting for the generalized Einstein
relation. In Sec.~\ref{sec:iteration}, we derive the appropriately
generalized Gummel iteration for the model. Numerical data for three
example calculations are given in Sec.~\ref{sec:results}, comparing
the proposed method with that of Ref.~\onlinecite{Knapp2010}. Finally,
we draw our conclusions in Sec.~\ref{sec:conclusions}. The procedure
is written for hole-transporting and bipolar devices in
Appendix~\ref{sec:app_bipolar}, and the scaling factors we use are
tabulated in Appendix~\ref{sec:app_scaling}.  Throughout this article,
we use the symbol $e$ for the elementary charge.

\section{Mathematical model\label{sec:model}}
The starting point for the derivations of most macroscopic charge
transport models in semiconductors is the drift-diffusion-Poisson
system introduced for inorganic crystalline semiconductors by van
Roosbroeck,\cite{Roosbroeck1950}
\begin{align}
-\nabla \cdot J_n &= \partder{\rho_n}{t}{}\;,\label{eq:Roos1}\\
-\nabla \cdot J_p &= \partder{\rho_p}{t}{}\;,\label{eq:Roos2}\\
\epsilon_0\epsilon_r\Delta \psi &= e(\rho_n-\rho_p-C)\;,\label{eq:Roos3}\\
J_{\nicefrac{n}{p}} &=\pm \mu\rho_{\nicefrac{n}{p}}\nabla\psi - D\nabla\rho_{\nicefrac{n}{p}}\;.\label{eq:Roos4}
\end{align}
Here, $J$ is the particle current, $\rho$ is the density, $\psi$ is
the electrostatic potential, $C$ is the dopant density (assumed to be
positive in case of $n$-type doping), $\mu$ is the electron mobility,
and $D$ is the diffusion constant. The subscripts $n$ and $p$ refer to
electrons and holes, respectively, and the total charge current is
given by $-e(J_n-J_p)$. To simplify the notation, we write the
equations henceforth only for electrons using the electron density
$n=\rho_n$. The following derivation is equally applicable for holes, since the equations in this case only differ by the sign of the drift term in the
continuity equation and the carrier density in the Poisson equation.
Appendix~\ref{sec:app_bipolar} gives the extension of our results to
bipolar and hole-transport devices.

The model of Eqs.~\ref{eq:Roos1}--\ref{eq:Roos4} was derived for
crystalline silicon-based semiconductors\cite{Juengel2009} and has
been applied in that area for many years with great success. Although
there is no \textit{ab initio} derivation, application of
Eqs.~\ref{eq:Roos1}--\ref{eq:Roos4} to organic semiconductors seems
promising since the charge continuity and Poisson equations
intuitively apply in some form. Similar models are used for organic
electronics in, \textit{e.g.}, Refs.~\onlinecite{Knapp2010,Xu2011}.

In fact, Eqs.~\ref{eq:Roos1}--\ref{eq:Roos4} must be extended in order
to accurately describe disorded organic
semiconductors.\cite{Scher1991} Available models include the extended
Gaussian disorder model (EGDM) and the extended correlated disorder
model (ECDM).\cite{Coehoorn2005,Bouhassoune2009} Both models assume a
hopping mechanism for charge transport with a Gaussian density of
states. The EGDM and ECDM introduce density- and field-dependence into
the mobility and diffusion coefficients $\mu(T,\rho,\der{\psi}{x}{})$
and $D(T,\rho,\der{\psi}{x}{})$, respectively. The methods that we
present in Secs.~\ref{sec:scaling}, \ref{sec:discretization} and
\ref{sec:iteration} are applicable to arbitrary $\mu$ and $D$; here we
discuss the EGDM and ECDM since they are recent and popular models.

It should be noted that in the context of so-called deep traps,
introduced \textit{e.g.} in Refs. \onlinecite{Knapp2010,Schober2011},
the trapped carriers act as a stationary background charge and their
treatment is completely analogous to the doping. Our method should
therefore also lead to better performance in situations where deep
traps are considered.

We restrict the discussion to one-dimensional steady-state current
simulations. It is useful to introduce the functions $F_1$ and $F_2$,
writing the steady-state system as
\begin{align}
F_1(n,\psi)&=\derop{x}{}\left[ - n\mu\left(T,n,\der{\psi}{x}{}\right) \der{\psi}{x}{} \right.
\nonumber \\
&+ \left. D\left(T,n,\der{\psi}{x}{}\right)\der{n}{x}{} \right] =0\;,  \label{eq:ddegdm1} \\
F_2(\rho,\psi)&=\epsilon_0\epsilon_r\der{\psi}{x}{2} - e(n - C) =0\;,\label{eq:ddegdm2}
\end{align} 
where $T$ is the temperature. When the system is accurately described
by Maxwell-Boltzmann statistics, the classical Einstein relation
holds,
\begin{align}
D=V_\text{th} \mu\;, \label{eq:er}
\end{align}
with the thermal voltage $V_\text{th} = k_BT/e$. Note that in this
equation the Boltzmann constant $k_B$ is taken to have units
$\mathrm{JK}^{-1}$. In order to align notation with other authors in
the field, we will frequently use $k_B$ in units of
$\mathrm{eVK}^{-1}$. To still be able distinguish these two
conventions we will use $k$ in this case.

When facing high charge-carrier densities, the effects of the Pauli exclusion
principle become important and one is forced to resort to Fermi-Dirac
statistics. In the context of organic semiconductors, it was first
pointed out by Roichmann and Tessler\cite{Roichman2002} that one
should then use a generalized version of Eq.~\ref{eq:er},
\begin{align}
D=\frac{n\mu}{e}\left(
\partder{n}{\varphi_{n}}{}\right)^{-1} \;, \label{eq:ger}
\end{align}
where $\varphi_n$ is the electron quasi-Fermi level (sometimes
referred to as the chemical potential; we use the term quasi-Fermi
level to avoid confusion and to stress that it is a non-equilibrium
quantity). Within the EGDM/ECDM framework, Eq.~\ref{eq:ger} is usually written as
\begin{align}
D=g_3V_{\text{th}}\mu\;,\qquad g_3=\frac{1}{kT}n\left(
\partder{n}{\varphi_{n}}{}\right)^{-1} \;,
\end{align} 
and $g_3$ is termed the diffusion enhancement. We adopt this
convention for clarity and compatibility with other authors. 

For a given density, the quasi-Fermi level is defined implicitly by the identity
\begin{align}
  n(\varphi_{n}) = \int_{-\infty}^\infty \mathrm{DOS}(E) \left[ 1
    +\exponential{\frac{E-\varphi_{n}}{kT}}\right ]^{-1}
  \mathrm{d}E\;, \label{eq:gfint1}
\end{align}
where DOS($E$) represents the density of states.

Equation~\ref{eq:ger} is called the generalized Einstein
relation\cite{Ashcroft1976} and accounts for the fact that charge
carriers near the quasi-Fermi level are more likely to contribute to
diffusion. The derivative occurring in Eq.~\ref{eq:ger} can be
computed analytically by differentiating Eq.~\ref{eq:gfint1}. The
differentiated expression can be found in \textit{e.g.}, the Appendix
of Ref.~\onlinecite{Mensfoort2008}.

\section{Scaling\label{sec:scaling}}
In SI units, the absolute values of the carrier density $[\mathrm{m}^{-3}]$ and
electrostatic potential $[\mathrm{V}]$ can differ by over $20$ orders of
magnitude. To avoid numerical problems we thus rescale the
equations. Several scaling schemes have been proposed by other authors
for both inorganic crystalline and organic
devices.\cite{Brezzi2005,Bonham1977} We find that the scaling leading
to the most stable behavior is state-dependent.

We suggest scaling the density by its maximum value (taking doping
into account). We refer to the density scaling factor as
$n_\mathrm{scal}$. Since we use a thermionic injection
model~\cite{Scott1999}, which can be coupled with barrier
lowering~\cite{Emtage1966}, the boundary values and therefore maxima
of the density are not known a priori. The scaling is therefore
applied at every iteration. The potential is scaled by
$\psi_\mathrm{scal}$, the maximum value of (the absolute value of) the
applied voltage $V_\mathrm{appl}$ and the thermal voltage. This
ensures that the scaled potential is reasonably small even for high
applied voltages. On the other hand, we force the scaling parameter to
be at least the thermal voltage to avoid scaling by zero. Since the
mobility may differ strongly between materials, we scale all
mobilities by their zero-field, zero-density limit $\mu_0$. Finally,
space is scaled by $L$, the total thickness of the device. The full
list of the scaling parameters including the resulting scaling of the
current density is given in Table~\ref{tab:scaling}.

The scaled system of equations reads (using the same symbols for the
scaled quantities)
\begin{align}
  F_1(\rho,\psi)&= \derop{x}{}\left[D\left(T,n,\der{\psi}{x}{}\right)\right.\nonumber \\
  &\times \left. \left(
      \frac{p_r n}{g_3(T,n)}\der{\psi}{x}{} - \der{n}{x}{}
    \right)\right]
  =0\;,    \label{eq:implicit1}\\
  F_2(n,\psi)&=\der{\psi}{x}{2} - \lambda^{-2}\left(n - C\right) =0\;, \label{eq:implicit2}
\end{align}
where we have used the constants
\begin{align}
p_r:=\frac{\psi_\text{scal}}{V_\text{{th}}}\;,\qquad
\frac{1}{\lambda^2}:=\frac{L^2en_\text{scal}}{\varepsilon\varepsilon_0\psi_\text{scal}}\;.
\end{align}
Here, $\lambda$ is the scaled Debye length.

This scaling ensures that the potential and density vary between $0$
and $1$ for most physical situations. Failing to implement a suitable
scaling scheme can result in numerical instability and loss of
accuracy. Furthermore, it reduces the number of constants used in the
actual computations. Except where explicitly stated, we henceforth use
the scaled quantities.

\section{Discretization\label{sec:discretization}}
To solve Eqs.~\ref{eq:ddegdm1} and \ref{eq:ddegdm2} (note that these are the unscaled equations), Scharfetter
and Gummel\cite{Scharfetter1969} proposed a scheme for the case of
constant mobility and diffusion. Instead of using central or upwind
differences to approximate the current, a weighted difference
depending on the field is used. It can be derived as a solution of the
boundary value problem (BVP)
\begin{align}
\begin{aligned}
  J_{i+\half} &= \mu \left(n\nabla\psi
    - V_\text{th} \nabla n\right)\;,\\
  n(x=x_i)&=n_{i}\;,\\
  n(x=x_{i+1})&=n_{i+1}\;,\end{aligned} \label{eq:sg_ode}
\end{align}
which describes the physics on a sub-interval for given carrier
densities at the boundaries. The BVP of Eq.~\ref{eq:sg_ode} is
first order, but we can prescribe two boundary values because
$J_{i+\half}$ is a free parameter of the system, and there is only one
value of $J_{i+\half}$ that admits a solution. This value can be used
as an approximation of the current, assuming that the mobility,
diffusion coefficient and electric field are locally constant. Expressed in terms of $n$ and
$\psi$, it reads
\begin{align}
\begin{aligned}
  J_{i+\half}&= \mu \left(\frac{ \psi_{i+1}
      -\psi_i}{h_i}\right) \frac{n_{i+1} - \exp
    \left[\frac{\psi_{i+1}-\psi_i}{V_{\text{th}}} \right]n_{i}}{1- \exp
    \left[\frac{\psi_{i+1}-\psi_i}{V_{\text{th}}}
    \right]}.\end{aligned} \label{eq:sg}
\end{align}
The scheme can be viewed as an upwind scheme, where the amount of
upwind difference depends on the local strength of the field. This is
easily seen by considering the asymptotic behavior of $J_{i+\half}$ when the
field is large or small,
\begin{align}
 J_{i+\half} \sim -\mu V_{\text{th}} \frac{n_{i+1}-n_{i}}{h_i},\quad\text{as }\partder{\psi}{x}{}\rightarrow 0\;,
\end{align}
so that for small fields we use a classic central difference
approximation for the gradient of the density, \textit{i.e.}, the
diffusive part of the current. When the magnitude of the field is
large, we obtain
\begin{align}
 J_{i+\half} \sim 
 \mu  \frac{\psi_{i+1}-\psi_{i}}{h_i} n_{i},\quad\text{as }\partder{\psi}{x}{}\rightarrow \infty\;,
\end{align}
and
\begin{align}
 J_{i+\half} \sim 
 \mu  \frac{\psi_{i+1}-\psi_{i}}{h_i} n_{i+1},\quad\text{as }\partder{\psi}{x}{}\rightarrow -\infty\;,
\end{align}
which are exactly upwind approximations for the density.

Upwind schemes introduce artificial diffusion into the
solution,\cite{Smith1985} which is a source of error. In the
case of constant mobility and the simple Einstein relation of
Eq.~\ref{eq:er}, it can be proven that the discretization introduced
in Eq.~\ref{eq:sg} yields the optimal artificial diffusion
(\textit{i.e.}, the minimum required for numerical stability). \cite{Doolan1980}

Within high-density capable models such as EGDM/ECDM, the generalized Einstein relation of
Eq.~\ref{eq:ger} increases physical diffusion. Use of Eq.~\ref{eq:sg}
to approximate the current is then suboptimal in terms of the accuracy
of the solution because more artificial diffusion is added than is
required for stability. Using the generalized Einstein relation
Eq.~\ref{eq:ger} and applying the scaling introduced in Sec.~\ref{sec:scaling}, we obtain the following generalization of the BVP of
Eq.~\ref{eq:sg_ode},
\begin{align}
\begin{aligned}
  J_{i+\half} &=J_\text{scal} \mu_{i+\half} \left( n\nabla\psi
    -\frac{g_{3,i+\half}}{p_r}
    \nabla n\right)\;, \\
  n(x=x_i)&=n_{i}\;,\quad
  n(x=x_{i+1})=n_{i+1}\;.\end{aligned} \label{eq:sg_ode2}
\end{align}
This corresponds to the current in Eq.~\ref{eq:implicit1}.

Assuming $\mu$, $g_{3}$ and $\nabla\psi$ to be constant on each interval, or put equivalently, approximating them by their value at $i+\half$, we solve
Eq.~\ref{eq:sg_ode2} as before to get a properly-adjusted
discretization scheme,
\begin{align}\begin{aligned}
    J_{i+\half} &\approx J_\text{scal}  \mu_{i+\half} \frac{\psi_{i+1} -\psi_i}{h_i}\\
    &\times \frac{n_{i+1} - \exp \left[
       \frac{p_r}{g_{3,i+\half}}\left(\psi_{i+1}-\psi_i\right) \right]n_{i}}{1-
      \exp \left[ \frac{p_r}{g_{3,i+\half}}\left(\psi_{i+1}-\psi_i\right)
      \right]}\;.\end{aligned}\label{eq:sg_scal}
\end{align}
This scheme accounts for the additional physical diffusion in our
system by decreasing the artificial diffusion, improving the accuracy
of the current.

In the EGDM/ECDM, $\mu=\mu_0(T)g_1(n,T)g_2(\nabla\psi,T)$. In this
case $g_1$ and $g_3$ at half-integer gridpoints can be obtained by
various averaging methods. Throughout this work we use the geometric
average
\begin{align}
g_{\ast,i+\half}\approx \sqrt{g_{\ast,i+1}g_{\ast,i}}\;,
\end{align}
which is an exact interpolant for exponentially varying
functions. Note that since $g_2$ depends on the gradient of the
potential, it is already known on half-integer gridpoints only.

\section{Iterative solution of the discretized
  system\label{sec:iteration}}
To solve the system of discretized equations, Eqs.~\ref{eq:ddegdm1}
and \ref{eq:ddegdm2}, using the discretization of
Sec.~\ref{sec:discretization}, one can either use the Newton algorithm
or the so-called Gummel iteration, a physically-motivated variant of
the Gauss-Seidel relaxation scheme.\cite{Gummel1964}

Numerical comparisons as in the work of Knapp \textit{et
  al}.\cite{Knapp2011} and additional theoretical considerations show
that Newton's method converges in fewer iterations than the Gummel
method. However, Newton iterations are generally more expensive to
compute. Furthermore, Newton's method often requires initial values in
the vicinity of the solution in order to obtain
convergence. 

This is a severe problem for optimization and parameter extraction,
where the solver is routinely called several hundred times. In these
cases a single failure to converge can be catastrophic for the
optimizer. We therefore derive an iterative solution scheme in the
spirit of Gummel.\cite{Gummel1964} As a starting point we will use the
continuous (non-discretized) problem to avoid working in spaces of
grid functions.

The idea is to alternately solve $F_1$ for $n$ at a given $\psi$ and
$F_2$ for $\psi$ and given $n$ until self-consistency is achieved.

Since the drift-diffusion-Poisson system of equations can be
nonlinear (even when $\mu$ and $D$ are constant; there is implicit
dependency of \textit{e.g.}, $\psi$ on $n$), we must linearize first. Our iteration starts with the Poisson equation.

\begin{figure}
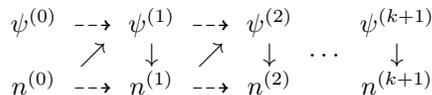

\centering
\[\begin{array}{ccccccc}
\psi^{(0)} & \dashrightarrow & \psi^{(1)} & \dashrightarrow & \psi^{(2)} &  & \psi^{(k+1)}\\
 & \nearrow & \downarrow & \nearrow & \downarrow & \cdots & \downarrow\\
n^{(0)} & \dashrightarrow & n^{(1)} & \dashrightarrow & n^{(2)} &  & n^{(k+1)}\nonumber
\end{array}\]
\caption{\label{fig:iteration} The arrows $\nearrow$ and $\downarrow$ represent solution of the
Poisson equation and solution of the steady-state continuity equation,
respectively. Dashed arrows show a quantity being carried forward for
use in the next iteration.}
\end{figure}

\subsection{Linearization of the Poisson equation} 
The essence of the Gummel iteration is to consider the implicit
nonlinearity of the Poisson equation explicitly, i.e.,
\begin{align}
F_2(n,\psi)&=\der{\psi}{x}{2} -  \lambda^{-2}\left(n(\psi) - C\right) =0\;.
\end{align}
In the original derivation, the nonlinearity of $n$ in $\psi$ is
revealed by the explicit use of Maxwell-Boltzmann statistics. This
does not apply in our case, and we must use Eq.~\ref{eq:gfint1}
instead. However, since even in Eq.~\ref{eq:gfint1} the dependence on
$\psi$ is not explicit, we must first introduce another quantity, the
quasielectrochemical potential $\zeta_{n}$, as defined in
\textit{e.g.}, Ref.~\onlinecite{Woellner2011},
\begin{align}
\zeta_{n} := \varphi_{n} - q\psi\;.\label{eq:qep}
\end{align}
The quantity $\zeta_n$ can be interpreted as the single driving
potential for the charge carrier transport. One must be aware that
there are several conventions in defining these levels, which are not
consistent with each other. We use the distinction between the
different levels as introduced in
Ref.~\onlinecite{Woellner2011}. However, we use a definition of the
quasi-Fermi level consistent with Ref.~\onlinecite{Mensfoort2008},
since the $g_3$-prefactor that we use in our computations is
introduced there. Therefore, the quasi-Fermi level and
quasielectrochemical potential are switched with respect to those of
Ref.~\onlinecite{Woellner2011}. The difference stems from the choice of
reference point with respect to which we define the DOS.  We need
$\zeta_n$ here to reformulate the Gauss-Fermi integral of
Eq.~\ref{eq:gfint1}. Since this equation uses an unscaled $\psi$ we
need to unscale our variable when we put it in
\begin{align}\begin{aligned}
    n(\zeta_{n},\psi) &=
    \frac{N_{\text{sites},n}}{n_\text{scal}\sqrt{2\pi}}
    \int_{-\infty}^\infty \exponential{-\frac{E^2}{\sqrt{2}\sigma}} \\
    &\times \left[{1+\exponential{\frac{E - \zeta_{n} -
            q\psi_\text{scal}{\psi}}{k
            T}}}\right]^{-1}\mathrm{d}E\;,\end{aligned}\label{eq:gfint2}
\end{align}
where $N_{\mathrm{sites}}$ is the site density. We deliberately use
$q$ here instead of $e$ because its numerical value is $1$. This can
be understood from the consideration that all energies in the
numerator of the exponential in Eq.~\ref{eq:gfint2} are in units of
$\mathrm{eV}$, whereas the product $e\psi_\text{scal}\psi$ has unit
$\mathrm{J}$. These units differ by the numerical value of the
elementary charge $e$, hence $q$ in Eq.~\ref{eq:qep} is $1\mathrm{C}$
and can be omitted in numerical calculations.

We find it useful to define the 'Poisson equation operator'
\begin{align}\begin{aligned}
    \operatorname{Poiss}\left(\psi,n(\psi)\right) &:=\Delta \psi -
    \lambda^{-2}(n(\psi)-C)\;,
\end{aligned}\label{eq:poissop}
\end{align}
which we wish to linearize at the current iterate
$(\psi^{(k)},n^{(k)})$, where $k$ denotes iteration number. To avoid
unnecessarily lengthy formulas, we introduce the notation
\begin{align}
\partder{X}{Y}{}\on{(\psi^{(k)},n^{(k)})} :=\partder{X}{Y}{}\on{k}\;.
\end{align}
The linearization is performed using functional
derivatives,\cite{Courant2007}
\begin{align}
\begin{aligned}
  &\operatorname{Poiss}\left(\psi,n(\psi)\right) \\ &\approx
    \underbrace{\Delta \left(\psi^{(k)}\right)-\lambda^{-2}\left(n^{(k)}-C \right)}_{\operatorname{Poiss}\left(\psi^{(k)},n^{(k)}\right)}\\
    &+\underbrace{\left[-\frac{1}{\lambda^2}\partder{n}{\psi}{}\on{k}
        +\Delta\right]}_{\partder{\operatorname{Poiss}}{\psi}{}\on{k}}\left(\psi-\psi^{(k)}\right)\;.
\end{aligned} \label{eq:linpoiss}
\end{align}
The term $\partder{n}{\psi}{}\on{k}$ is understood as a (pointwise)
multiplication operator.  Since the central difference approximation
of the Laplacian and the pointwise approximation of
$\partder{n}{\psi}{}\on{k}$ are linear and continuous as projections
onto finite dimensional spaces, the discretization commutes with the
linearization.

The derivative $\partder{n}{\psi}{}\on{k}$ is
straightforwardly related to the $g_3$ factor. Comparing
Eqs.~\ref{eq:gfint1} and \ref{eq:gfint2}, we get
\begin{align}
  \partder{n}{\psi}{}\on{k}= q\psi_\text{scal} \partder{n}{\varphi_{n}}{}\on{k}\;.\label{eq:dndpsianalogue}
\end{align}
We can rewrite the derivative with respect to $\varphi_n$,
\begin{align}
  \partder{n}{\varphi_{n}}{}=\frac{1}{kT g_{3}}n\;,
\end{align}
which leads to
\begin{align}
\partder{n}{\psi}{}\on{k}&=\frac{p_r}{g_3(n^{(k)})}n^{(k)}\;.\label{eq:dndpsi}
\end{align}
Equation~\ref{eq:dndpsi} is highly convenient, since $g_3$ and $n$ are
computed elsewhere and can be reused here at virtually no cost. The
linearization process thus doesn't require computation of any
additional quantities.

Using Eq.~\ref{eq:dndpsi}, we get for the derivative of the
Poisson-operator
\begin{align}
  \partder{\operatorname{Poiss}}{\psi}{}\on{k} = \left[ -
   \frac{p_r}{\lambda^2}\frac{n^{(k)}}{g_{3}(n^{(k)})}+\Delta \right]\;.
\end{align}
To complete the linearization, we insert this result into
Eq.~\ref{eq:linpoiss}, set Poiss$(\psi,n(\psi))=0$ and
$\psi=\psi^{(k+1)}$, and solve for $\psi^{(k+1)}$. This yields

\begin{align}
\psi^{(k+1)}&=\left(
\Delta-\frac{p_r}{\lambda^2}\frac{ n^{(k)}}{g_3(n^{(k)})}
\right)^{-1}
\left[
\frac{1}{\lambda^2}\left(n^{(k)}-C\right)\phantom{\frac{ n^{(k)}}{g_3(n^{(k)})}}\right.\nonumber\\
&\left. -\frac{p_r}{\lambda^2}\frac{ n^{(k)}}{g_3(n^{(k)})}\psi^{(k)}
\right]
\;.
\label{eq:psiit}
\end{align}
We now switch back to the discretized system. As mentioned before, the
order in which we linearize and discretize does not matter. We
discretize using the following expressions,
\begin{align}
\frac{p_r}{\lambda^2}\frac{n^{(k)}}{g^{(k)}_{3}} &=\frac{p_{r}}{\lambda^2}\frac{n_i^{(k)}}{g_{3,i}^{(k)}}\;,\\
\Delta &\approx \left(
\begin{array}{ccccccc}
1 &  &  &  &  &  & \\
1 & -2 & 1\\
 & 1 & -2 & 1\\
 &  & \ddots & \ddots & \ddots\\
 &  &  & 1 & -2 & 1\\
 &  &  &  & 1 & -2 & 1\\
 &  &  &  &  &  & 1
\end{array}
\right)\;,\label{eq:poissmat}
\end{align}

  \begin{figure*}
  \centering
  \includegraphics{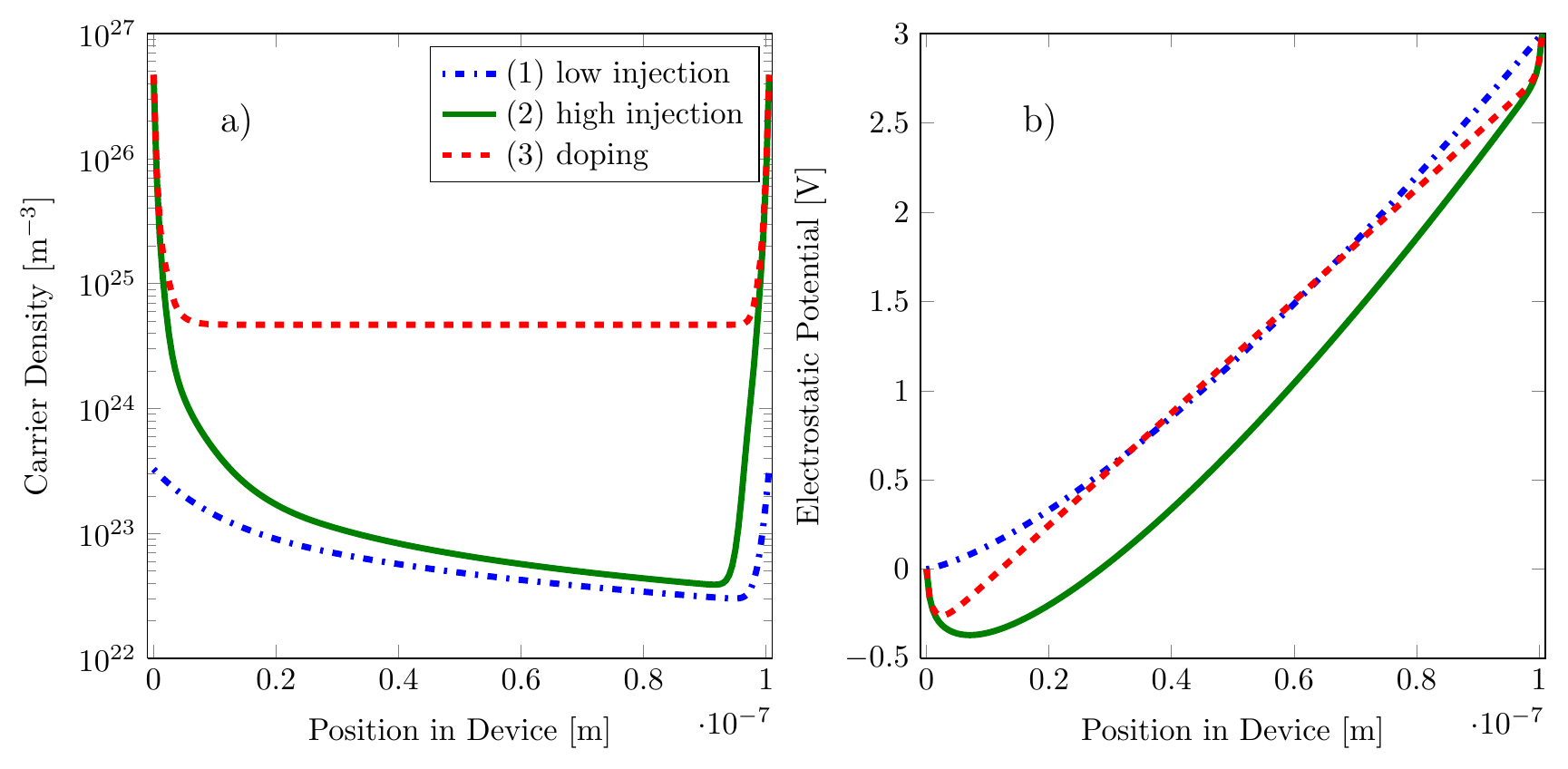}
  \caption{(Color online) Simulation results with the proposed
    algorithm. Where the method of Knapp \textit{et al}.\ (not shown)
    converges, the results coincide with ours. The applied voltage in
    the simulation was set to $3\mathrm{V}$. Model parameters are
    given in Table \ref{tab:para}.}\label{fig:simulation}
  \end{figure*}
and solve for $\psi^{(k+1)}$, setting the first and the last (spatial
grid) values of the quantity in the square brackets of
Eq.~\ref{eq:psiit} to the boundary values. The boundary values of
$\psi$ and $n$ are determined by the applied voltage and the charge
injection, respectively. The matrix elements not shown in
Eq.~\ref{eq:poissmat} are zero.

 Equation \ref{eq:psiit}, which we derived by linearizing the poisson
 operator, properly generalizes the Gummel iteration scheme to the
 case of a generalized Einstein relation.

\subsection{Linearized Continuity equation}
Since it is comparatively straightforward, we give the linearization
for the continuity equation on the discretized level. We do not
consider the implicit nonlinearity via $\psi(n)$ here, and linearize
the mobility by assuming that $\mu^{(k+1)}$, the mobility at the
$k+1$-th iteration, depends on $n^{(k)}$ and not $n^{(k+1)}$. Before
giving the full expression for the linearized and discretized current,
it is convenient to introduce the function
\begin{align}
G^{(k)}_{n,i} : =  \exp
    \left[\frac{p_r}{g_{3,i+\half}(n^{(k)})}\left(\psi_{i+1}^{(k+1)}-\psi_i^{(k+1)}\right)
    \right]\;.
\end{align} 
Substituting this definition into Eq.~\ref{eq:sg_scal}, the
discretized and linearized current is given by
\begin{align}
  J_{i+\half}^{(k+1)} &\approx J_\text{scal}\mu_{i+\half}\left(n^{(k)},\psi^{(k+1)}\right)
   \left(\frac{\psi_{i+1}^{(k+1)} -\psi_i^{(k+1)}}{h_i}\right)\nonumber \\
  &\times
  \left(\frac{n_{i+1}^{(k+1)} - G^{(k)}_{n,i}n_{i}^{(k+1)}} {1-
    G^{(k)}_{n,i}}\right)\;. \label{eq:fulldisccont}
\end{align}

This expression is linear in $n^{(k+1)}$. Therefore the central
approximation of the continuity equation $D^0 J_{n}^{(k+1)}=0$, which
is what is solved in practice, is clearly also linear in
$n^{(k+1)}$. Equations \ref{eq:psiit}--\ref{eq:fulldisccont} define
the discretized and linearized system and are accessible to direct
solution methods.

A diagram illustrating the iteration scheme resulting from the linearizations presented in the preceding sections can be found in Fig.~\ref{fig:iteration}.

\section{Simulation results\label{sec:results}}
To highlight the advantages of the schemes described above, we have
simulated a simple single-layer structure using the ECDM model. We
consider three different situations: First, we simulate a low density
situation without doping --- here we show that our approach reproduces
known results with a small gain in computational efficiency. Second,
we simulate a device with high carrier densities due to a low
injection barrier --- here we observe a significant performance gain
with the proposed method. Finally, we simulate a strongly-doped
device, with approximately every $100$th molecule replaced by a
dopant. In this third case, conventional methods fail but our
procedure converges. The parameters used for the simulations are
collected in Table~\ref{tab:para}. The resulting charge carrier
densities and potentials at $3\mathrm{V}$ are shown in
Fig.~\ref{fig:simulation}. All of the device simulations, including
those using the conventional iteration, used the discretization
described in Sec.~\ref{sec:discretization} and had identical initial
values. Our calculations did not include the barrier-lowering effect
of image-charges in the metal contacts. Its inclusion would have the
effect of increasing the carrier injection, further enhancing the
computational advantage of our method.
\begin{table}
  \caption{Simulation parameters for the three example cases that we consider. Simulation results are given in Figs.~\ref{fig:simulation}--\ref{fig:gridcon}.} \label{tab:para}
\begin{tabular}{c|c|c|c}
  \hline\hline
  Parameter & \phantom{X}(1)  \phantom{X}&  \phantom{X}(2)  \phantom{X}&  \phantom{X}(3) \phantom{X} \\ \hline
  Electrode WF [eV] & -5  & -4.6  & -4.6  \\
  Doping intensity & $0\%$ &$0\%$ &$1\%$ \\
\hline
  Mobility model & \multicolumn{3}{c}{ECDM} \\
  Injection model & \multicolumn{3}{c}{Thermionic} \\
  LUMO [eV] & \multicolumn{3}{c} {-4.5}\\
  Site density [m$^{-3}$] & \multicolumn{3}{c} {$2\times 10^{27}$} \\
  DOS width [eV] & \multicolumn{3}{c} {0.13} \\
  Temperature [K] & \multicolumn{3}{c} {300} \\
  Device Length [nm] & \multicolumn{3}{c} {100} \\
  ECDM-C parameter & \multicolumn{3}{c} {0.29} \\
  $\mu_0$  [m$^{2}$(Vs)$^{-1} $] & \multicolumn{3}{c} {$4.5\times 10^{-6}$}\\
  \hline\hline
\end{tabular}
\end{table}

  \begin{figure*}
  \centering
  \includegraphics{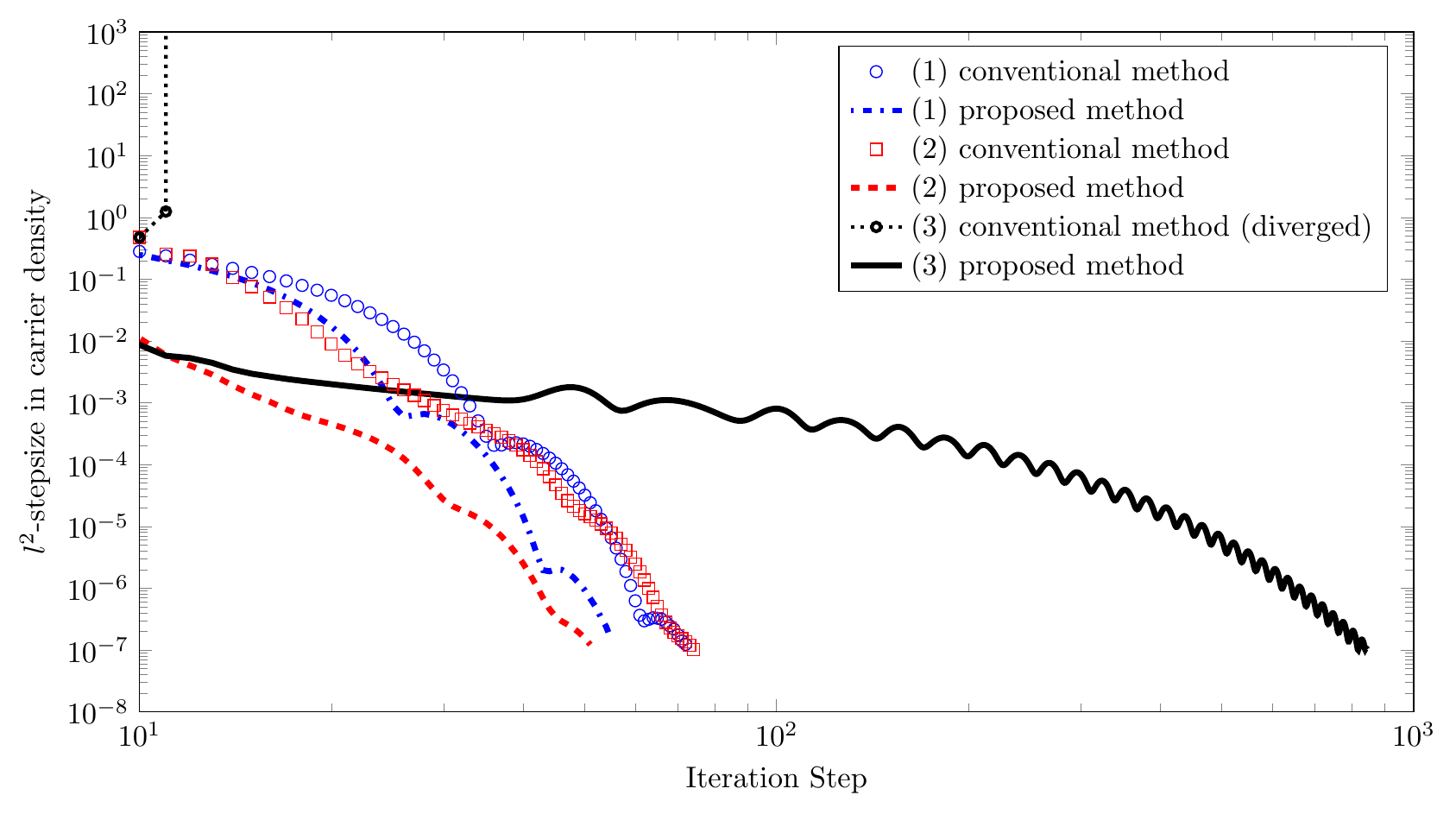}
  \caption{(Color online) Plot of convergence speed in $l^2$-stepsize of
    the carrier density for the three different scenarios given in Table
    ~\ref{tab:para}. The 'proposed' method is that described in this paper,
    while the 'conventional' method is the iteration given by Knapp
    \textit{et al.}\cite{Knapp2010} Parameter values are given in
    Table~\ref{tab:para}.}\label{fig:convplot}
  \end{figure*}

\subsection{Convergence}
In Fig.~\ref{fig:convplot}, we compare the behavior of the method
proposed in this paper, with the method presented by Knapp \textit{et
  al.},\cite{Knapp2010} which we call the 'conventional method'.

The main difference between the approaches lies in the linearization
of the Poisson equation. Knapp \textit{et al.} use the expression
(written in the framework of our scaling for the reader's convenience)
\begin{align}
\psi^{(k+1)}&=\left(
\Delta-\frac{p_r}{\lambda^2} n^{(k)}
\right)^{-1}
\left[
\frac{1}{\lambda^2}n^{(k)}-\frac{p_r}{\lambda^2} n^{(k)}\psi^{(k)}
\right]
\;,
\label{eq:psiitKnapp}
\end{align}
instead of Eq.~\ref{eq:psiit}. This differs from Eq.~\ref{eq:psiit} by
the absence of $g_3$ in the denominators and the doping profile, which
wasn't considered in Ref.~\onlinecite{Knapp2010}. The effect can be
viewed as a different damping. However, one should keep in mind that
our scheme is not derived from a perspective of damping but from the
use of functional derivatives for the Poisson operator introduced in
Eq.~\ref{eq:poissop}.

A similar difference can be seen in the discretization of the
continuity equation, where an additional $g_3$ turns up in the
denominator of the argument of the exponential function. As mentioned
above, this prevents the introduction of unnecessary artificial
diffusion.

In all three device simulations, the proposed method is superior to
the conventional method in terms of convergence speed, with the
difference being greater when the carrier densities are high. Our
convergence criterion was that the $l^2$-stepsize in the scaled charge
density fall below $10^{-7}$.

We find that the stability of the generalized iteration in
high-concentration regimes is greatly enhanced in comparison with the
conventional iteration. For example, the conventional approach fails
to converge for doping intensities $\geq 1\%$, whereas our method
converges for arbitrary doping intensities. The benefits of properly
incorporating diffusion into the numerical scheme could also be
important for the simulation of multilayer devices, and in particular
those with doped layers.

Even in the absence of doping, for large values of charge carrier
injection a speed-up of convergence is observed. In the low density
limit, the method blends into the one presented in
Ref.~\onlinecite{Knapp2010} because $g_3\rightarrow 1$, as
$n\rightarrow 0$.

\subsection{Grid convergence}
Since we introduced a scheme using upwind stabilization, it is
interesting to study the effect of varying gridpoint density on the
behavior of the solution. We have studied the effect of the number of
gridpoints on the current density for the three cases shown in
Table~\ref{tab:para}. We investigate uniform grids with up to $1000$
gridpoints.
  \begin{figure}
  \centering
      \includegraphics{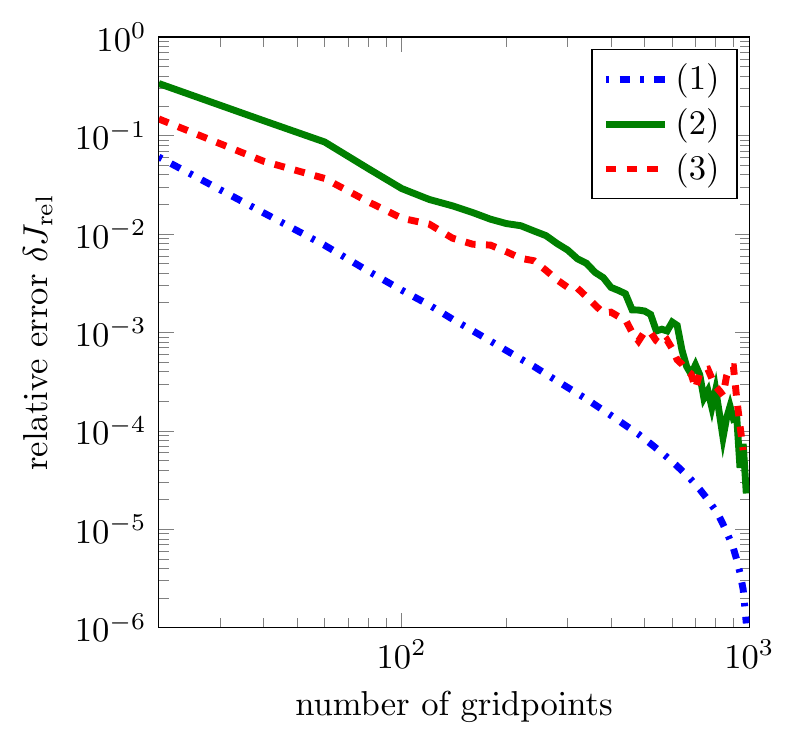}
      \caption{Convergence of the calculated current as the numerical
        grid is refined. Parameter values are given in
        Table~\ref{tab:para}. The relative error is defined as $\delta
        J_{\mathrm{rel}}=(J_N-J_{10^3})/J_{10^3}$, where $J_N$ is the
        calculated current with $N$ grid points.} \label{fig:gridcon}
  \end{figure}
  As shown in Fig.~\ref{fig:gridcon}, we observe convergence using the
  proposed method in all three cases. However, for high carrier
  densities the relative error in the current oscillates slightly with
  varying grid resolution. This is likely to be a result of the
  averaging procedure used on the mobility prefactors $g_{1-3}$, and
  is a topic for further investigation. The oscillations are, however,
  smaller than the general trend of convergence.

\section{Summary and Conclusions\label{sec:conclusions}} 
We have introduced a numerical method consisting of a scaling
scheme, a discretization scheme, and a fixed point iteration for the
van Roosbroeck drift-diffusion-Poisson system with arbitrary density- and
field-dependent mobility functions. The method properly accounts for
the implications of Fermi-Dirac statistics by incorporating the
generalized Einstein relation. The iteration is derived by linearizing the
Poisson equation with explicit treatment of the nonlinear dependence
of the density on the potential.

Failure to take consequences of the generalized Einstein relation into
account in the iteration results in numerical instability when the
density is large, \textit{e.g.}, in the case of strong doping or large
trap density, whereas the generalized method converges for arbitrary
doping. In other cases, we find that the generalized method is
significantly more efficient than the conventional method while
reproducing converged current values.

\begin{acknowledgments}
  We thank Ren\'e Pinnau from TU Kaiserslautern for fruitful
  discussions. C.K.F.~Weiler acknowledges funding from the BMBF
  project PARAPLUE, project number 03MS649A.
\end{acknowledgments}

\appendix
\section{Numerical Scheme for positive carriers\label{sec:app_bipolar}}
While the modelling details of bipolar devices (charge
generation/recombination) are outside the scope of this paper, we show
how to extend the method presented above to hole-transporting or bipolar
devices. For the iterative procedure, we include a hole continuity
equation, which has the same mathematical form as the electron
continuity equation, except the sign of the drift term is changed. As
for electrons, we use Scharfetter-Gummel for the discretization and
linearize the mobility by using the values of the last iteration. The
analogue of Eq.~\ref{eq:fulldisccont} for holes is

\begin{align}
  J_{p,i+\half}^{(k+1)} &\approx -J_\text{scal}
  \mu_{i+\half}\left(p^{(k)},\psi^{(k+1)}\right) \left(\frac{\psi_{i+1}^{(k+1)} -\psi_i^{(k+1)}}{h_i}\right)\nonumber \\
  &\times
  \left(\frac{p_i^{(k+1)} - G_{p,i}^{(k)}p_{i+1}^{(k+1)}} {1-
    G_{p,i}^{(k)}}\right)\;.
\end{align}
This allows us to solve the for the new iterate $p^{(k+1)}$ in the
same way as we did for the electron density in
Sec.~\ref{sec:iteration}. In the linearization of the Poisson
equation, we do not only need to linearize $n(\psi)$ but in an
analogous manner also $p(\psi)$. The effect of the potential on the
two carrier densities is the same except for a reversed sign,
\textit{i.e.,} the hole analogue of Eq.~\ref{eq:dndpsianalogue} has
a minus sign. However, since $n$ and $p$ appear in the poisson
equation with different signs, they appear in the linearized iteration
in the prefactor multiplying $\psi^{(k)}$ on the same footing
(\textit{i.e.}, $+\times-$ or $-\times +$). We obtain
\begin{align}
  \psi^{(k+1)}&= \partder{\operatorname{Poiss}}{\psi}{}\on{k}^{-1}\left[ \frac{1}{\lambda^2}\left(n^{(k)}-p^{(k)}-C \right) \phantom{\frac{p^{(k)}}{g_{3,p}^{(k)}}}\right.\nonumber\\
  &\left. - \left( \frac{p_r}{\lambda^2} \frac{n^{(k)}}{g_{3,n}^{(k)}}+\frac{p_r}{\lambda^2}
      \frac{p^{(k)}}{g_{3,p}^{(k)}} \right) \psi^{(k)}\right] \;.
\end{align}
Here, the linearized Poisson operator reads
\begin{align}
  \partder{\operatorname{Poiss}}{\psi}{}\on{k} = \left[ -\frac{p_r}{\lambda^2}\frac{n^{(k)}}{g_{3,n}^{(k)}}-
\frac{p_r}{\lambda^2}\frac{p^{(k)}}{g_{3,p}^{(k)}}+\Delta \right]\;.
\end{align}
The discretization and solution of the Poisson equation then works
exactly as described in Sec~\ref{sec:iteration}.

\section{Scaling factors\label{sec:app_scaling}}
Here we list the model scaling factors, which we use to force densities and potential to vary between zero and one. Note that the density and
current scaling factors have the potential to change during the
calculation, so should be updated after every iteration.
\begin{table}[h]
  \caption{Scaling factors for quantities considered in the model. The symbols in the second column are defined as the quantities in the third column.\label{tab:scaling}}
\begin{tabular*}{0.42\textwidth}{lcc}
  \hline\hline
  Variable  & \hspace{0.3cm} Symbol \hspace{0.3cm} & Scaling factor \\\hline
  Potential & $\psi_\mathrm{scal}$ &$\max\{|V_\text{appl}|,V_\text{th}\}$\\
  Density & $n_\mathrm{scal}$ & $\max\{\rho(x=0),\rho(x=L),$\\
 &&  $\max_{x\in[0,L]} C(x)\}$ \\
  Length &$L$& $L$ \\
  Mobility & $\mu_\mathrm{scal}$ & $\mu_0(T)$\\
  Current & $J_\text{scal}$ & $\mu_\text{scal}\psi_\text{scal}n_\text{scal}L^{-1}$\\
  \hline\hline
\end{tabular*}
\end{table}

\end{document}